\documentclass[aps, twocolumn, superscriptaddress]{revtex4} 
\usepackage[switch]{lineno}
\usepackage{amsmath,mathtools,amssymb}
\usepackage[utf8]{inputenc} 
\usepackage[T1]{fontenc}    
\usepackage{booktabs}       
\usepackage{empheq}
\usepackage[parfill]{parskip}
\usepackage{graphicx}
\usepackage[active]{srcltx}
\DeclareMathOperator{\EX}{\mathbb{E}}

\DeclarePairedDelimiter\floor{\lfloor}{\rfloor}

\begin{document}
\newcommand{\bi}{\begin{itemize}}
\newcommand{\ei}{\end{itemize}}
\newcommand{\be}{\begin{equation}}
\newcommand{\fe}{\end{equation}}
\setlength{\parindent}{0.5cm}
\newcommand{\argmin}{\arg\!\min}
\newcommand{\argmax}{\arg\!\max}


\title{KuraNet: Systems of Coupled Oscillators that Learn to Synchronize}

\author{Matthew Ricci}
\affiliation{Data Science Initiative, Brown University, Providence, RI 02906, USA}
\affiliation{Department of Cognitive, Linguistic and Psychological Sciences, Brown University, Providence, RI 02906, USA}

\author{Minju Jung}
\affiliation{Department of Cognitive, Linguistic and Psychological Sciences, Brown University, Providence, RI 02906, USA}

\author{Yuwei Zhang}
\affiliation{Department of Physics, Nankai University, Tianjin, China}

\author{Mathieu Chalvidal}
\affiliation{Artificial and Natural Intelligence Toulouse Institute, Universit\'{e} de Toulouse, Toulouse, France}

\author{Aneri Soni}
\affiliation{Department of Neuroscience,
Brown University, Providence, RI 02906, USA}

\author{Thomas Serre}
\affiliation{Department of Cognitive, Linguistic and Psychological Sciences, Brown University, Providence, RI 02906, USA}
\affiliation{Artificial and Natural Intelligence Toulouse Institute, Universit\'{e} de Toulouse, Toulouse, France}
\affiliation{Department of Computer Science, Brown University, Providence, RI 02906, USA}

\date{\today}

\begin{abstract}
Networks of coupled oscillators are some of the most studied objects in the theory of dynamical systems. Two important areas of current interest are the study of synchrony in highly disordered systems and the modeling of systems with adaptive network structures. Here, we present a single approach to both of these problems in the form of ``KuraNet'', a deep-learning-based system of coupled oscillators that can learn to synchronize across a distribution of disordered network conditions. The key feature of the model is the replacement of the traditionally static couplings with a coupling \emph{function} which can learn optimal interactions within heterogeneous oscillator populations. We apply our approach to the eponymous Kuramoto model and demonstrate how KuraNet can learn data-dependent coupling structures that promote either global or cluster synchrony. For example, we show how KuraNet can be used to empirically explore the conditions of global synchrony in analytically impenetrable models with disordered natural frequencies, external field strengths, and interaction delays. In a sequence of cluster synchrony experiments, we further show how KuraNet can function as a data classifier by synchronizing into coherent assemblies. In all cases, we show how KuraNet can generalize to both new data and new network scales, making it easy to work with small systems and form hypotheses about the thermodynamic limit. Our proposed learning-based approach is broadly applicable to arbitrary dynamical systems with wide-ranging relevance to modeling in physics and systems biology.
\end{abstract}


\maketitle

A central theme in the study of complex systems is the emergence of order from disorder, and few formalisms capture this spirit better than networks of coupled oscillators. These systems comprise a set of rotors, each of which is influenced by both its own intrinsic dynamic properties and the extrinsic influence of other rotors in the network. When the intrinsic and extrinsic factors in the network are appropriately balanced, synchrony emerges. That is, oscillators assume a common phase and begin to rotate with a common frequency. The ubiquity of synchronization in nature, from the spiking of neurons~\citep{Ursey1999, Salinas2001} to the onset of seasonal diseases~\citep{He2003}, from the entrainment laser arrays~\citep{Strogatz1993} to the release of energy during an earthquake~\citep{Vasudevan2015}, makes oscillatory networks a topic of major practical and theoretical interest in the study of nonlinear and complex systems (for a review of coupled oscillators in complex networks, see \cite{Rodrigues2016}).  

However, real-world networks are rarely static, and nature abounds with network topologies that evolve towards particular dynamic goals. These adaptive networks add a new wrinkle in the study of oscillatory systems: if the modeler can learn the conditions giving rise to synchrony, can networks be trained to do it on their own? Understanding this more general scenario is crucial to understanding adaptive networks in nature, from modeling changes in brain connectivity~\citep{Knoblauch2012}, to the evolution of flocking behavior~\citep{Ramos2019}, and even the emergence of consensus in societies~\citep{Hansen2020}. Moreover, a system which has learned to synchronize provides computational evidence about the conditions which give rise to this collective behavior, lending empirical aid to a theoretically challenging problem. 

This paper proposes a general framework, \emph{KuraNet}, for learning in oscillatory systems in the spirit of recent work at the intersection between machine learning and dynamical systems~\citep{Chen2018, Champion2019, chalvidal2021}. KuraNet replaces the standard fixed coupling matrix of the oscillator network with a coupling \emph{function} whose input data comprises random realizations of intrinsic oscillator features and whose output consists of data-dependent couplings. KuraNet's coupling function takes the form of a deep neural network whose parameters are adjusted by gradient descent. After training, the coupling function stores information about the conditions supporting synchrony \emph{in general} and not just for a particular sample of disordered features. Unlike previous methods for learning the topology of oscillatory networks \citep{Brede2008}, our system can generalize to both new oscillator features and network sizes. 

We focus on the Kuramoto model~\citep{Kuramoto1975,Strogatz2000}, the archetypical system of coupled oscillators, though our method can be equally applied to other types of oscillatory (and non-oscillatory) dynamics. After a brief description of the approach, we apply it to two general types of collective oscillatory behavior: global synchrony and cluster synchrony. In the first case, all phases cohere and oscillators spin at a common frequency. We will show how this technique can easily discover couplings which encourage phase coherence and how these learned functions can shed light on the conditions for synchrony in even highly disordered forms of the Kuramoto model. Cluster synchrony, on the other hand, occurs when oscillators split into separate factions which are internally coherent but mutually desynchronized. In contrast, earlier work~\citep{Wang2005, Schaub2016, Tiberi2018} on cluster synchrony has focused on the related but distinct case in which clusters behave independently of each other instead of maintaining strict inter-group separation. We show how the approach can be used to split populations into separate clusters based on class labels and how such a method can be used for the segmentation of high-resolution images. We will conclude with a summary of key theoretical and practical problems to which the framework might be applied in the future. 

\section{Background}
\label{sec:background}
An oscillator is a dynamical system with a stable, periodic limit cycle. Malkin~\citep{Malkin1949}, Winfree~\citep{Winfree1967}, and others showed that, under weak coupling conditions, $n$ oscillators, each associated to a node on the graph $G=(N,E)$, can be reduced to a system of phases, $\theta(t) \in \mathbb{T}^n$, on the $n$-torus. The most famous such ``phase-reduced'' system of oscillators is the Kuramoto model, which we consider in the following form:
\begin{equation}
    \dot{\theta}_i = \omega_i + \sum_{j = 1}^n K_{ij} \sin(\theta_j(t - \tau_j) - \theta_i(t)) + b_i\sin(\theta_i).
    \label{eq:km}
\end{equation}
The dynamics of an oscillator, $i$, are determined by both intrinsic and extrinsic factors. Here, the intrinsic factors comprise the \emph{node features}: natural frequencies, $\omega_i$, external field strengths, $b_i$, and transmission delays, $\tau_i$. The external influence on an oscillator is determined by interactions with other phases in the rest of the network, $\theta_j$ for $j \neq i$, via the matrix of weights or couplings, $K \in \mathbb{R}^{n \times n}$ associated to $G$. 

\begin{figure*}[t!]
\centering
\includegraphics[width=\textwidth]{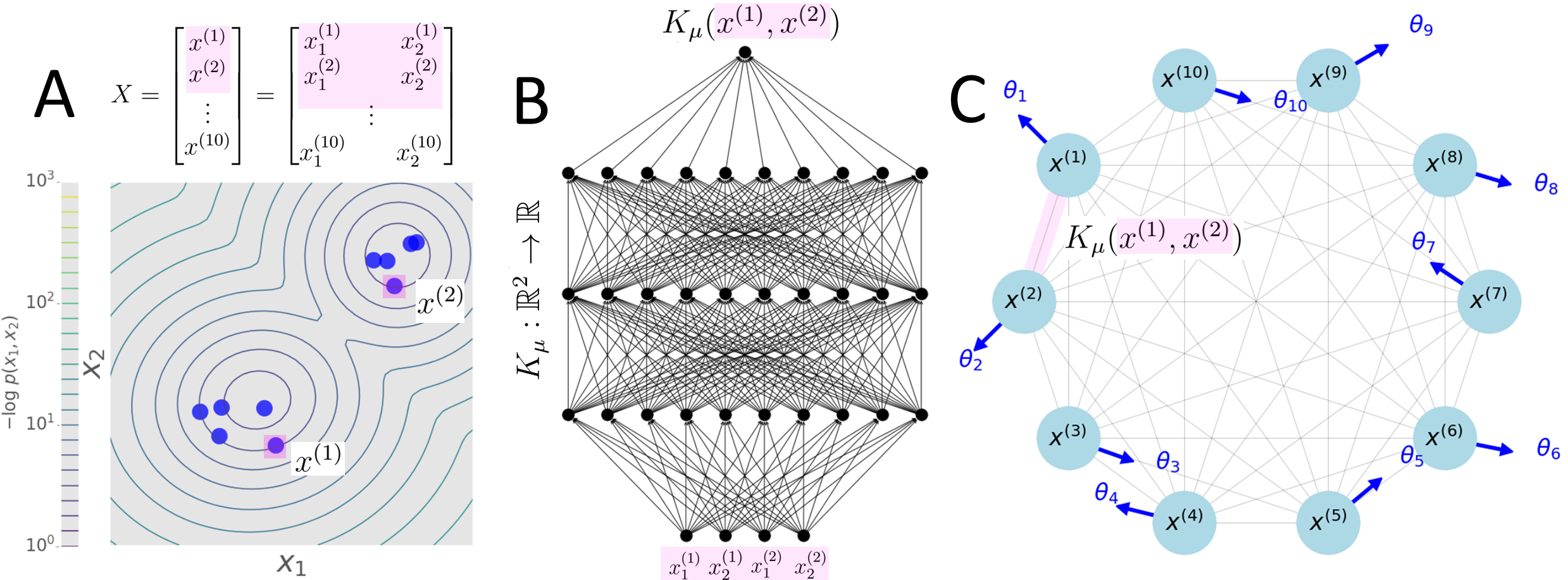}
\caption{\emph{KuraNet}. \emph{(A)}. First, a batch, $X\in \mathbb{R}^{n \times m}$, of random intrinsic oscillator features is sampled from the node feature space, $\mathcal{X}$. Here, we show a batch of $n=10$ oscillators each with feature, $x^{(i)} \in \mathbb{R}^2$, distributed according to a two-component gaussian mixture. Two oscillator features, $x^{(1)}$ and $x^{(2)}$, are shaded in pink and followed throughout KuraNet processing. \emph{(B)}. Then, all pairs $(x^{(i)}, x^{(j)})$ are passed in parallel (only one pair shown) through the coupling function, $K_{\mu}$, which takes the form of a deep neural network whose parameters, $\mu$, comprise synaptic weights and biases. \emph{(C)}. The output of the function on the pair of samples $(x^{(i)}, x^{(j)})$ becomes the coupling strength $K_{\mu}( x^{(i)}, x^{(j)})$ in a Kuramoto model, each of whose $i = 1, \ldots, n$ oscillators inherits the intrinsic feature $x^{(i)}$. A dynamics is simulated for the resulting Kuramoto model and evaluated according to a loss function whose output is used to adjust the parameters, $\mu$, of the coupling function.}
\label{fig:kuranet}
\end{figure*}

Disorder is introduced by placing probability distributions, $f, g$ and $h$, on $\omega_i$, $\tau_i$ and $b_i$, respectively. We write the joint distribution as $D$ and denote its sample space, the \emph{node feature space}, by $\mathcal{X}$. Global synchrony in the system is typically measured by the so-called global order parameter,
\begin{equation}
 Re^{i\phi} = \lim_{t \to \infty} \frac{1}{n}\sum_{j=1}^n e^{i \theta_j(t)},
\end{equation}
so that $R\approx 0$ when the oscillators are out of phase and $R = 1$ when oscillators are in phase ($i$ here denotes the imaginary unit, not an index). Below we will instead work with the circular variance, $V = 1 - R$.

The case of $g = h = \delta(0)\neq f$ with $f$ symmetric and unimodal has been extensively studied, especially for mean field couplings, $K_{ij} = \epsilon/n$, for $\epsilon > 0$ (see \cite{Acebron2005, Gupta2014} for excellent reviews). In this paradigmatic case, there exists a critical coupling, $\epsilon_c$, such that $V \to 0$ only when $\epsilon > \epsilon_c$. The critical coupling is inversely proportional to the variance of $f$ so that stronger coupling is required for more heterogeneous intrinsic frequencies. The case of more general disorder, the focus of the current study, has received less attention and disordered models are known to produce esoteric and poorly understood behaviors, like quasi-glassiness~\citep{Iatsenko2014a, Ottino2018}, hysteresis~\citep{Skardal2018} and multistability~\citep{Olmi2014}. 

\section{KuraNet} Various approaches to optimizing Kuramoto network topology have been proposed~\citep{Gleiser2006, Brede2008, Kelly2011}, though these techniques generally involve the optimization of a single coupling matrix for a given realization of random node features. Consequently, anything learned for this single instance does not necessarily generalize to networks with new disordered features or to networks of a different size. Such techniques are also limited in their ability to model adaptive oscillatory systems in the real world, which must presumably learn to synchronize in the presence of variable forms of disorder and not just a single example. Moreover, the pioneering approaches of \cite{Brede2008} and others typically rely on greedy, stochastic methods, whereby single network edges are turned on or off after each of many repetitions of the dynamics, making optimization prohibitively slow for large systems. 

Inspired by the growing body of research at the intersection of dynamical systems and machine learning~\citep{Nitzan2017, Champion2019, chalvidal2021}, we propose a new technique called \emph{KuraNet} for building oscillatory systems which can learn to synchronize (Figure~\ref{fig:kuranet}). The key difference between our approach and earlier techniques lies in our replacement of the traditional coupling matrix, $K$, with a differentiable coupling \emph{function}, $K_{\mu} : \mathcal{X}^2 \to \mathbb{R}$. This function maps from pairs of intrinsic node features to the coupling weight between those nodes. The coupling function takes the form of a deep neural network whose learnable parameters, $\mu$, are optimized to store knowledge about which couplings are appropriate for any realization of random node features. As a result, the coupling function can easily be transferred to new data. Moreover, since $K_{\mu}$ is applied to all pairs of nodes in parallel, the system can be automatically scaled up to much larger networks than those observed during training. 

KuraNet is trained by stochastic gradient descent on random batches of node features, $X\in \mathcal{X}^n$ (Figure~\ref{fig:kuranet}A). All $n(n-1)$ distinct pairs, $(x^{(i)}, x^{(j)})$, from a batch are passed in parallel through the coupling function, $K_{\mu} : \mathcal{X}^2 \to \mathbb{R}$ (Figure~\ref{fig:kuranet}B). The result is a data-dependent coupling matrix with elements $K_{\mu}( x^{(i)}, x^{(j)})$ which then participates in a Kuramoto dynamics (Figure~\ref{fig:kuranet}C) in which the $i^{\text{th}}$ oscillator inherits the random feature $x^{(i)}$ (e.g., $x^{(i)} = (\omega_i, b_i, \tau_i)$). Given an initial condition, $\theta_0$, Eq.~\ref{eq:km} is then solved with any off-the-shelf numerical method, and the solution $\theta(t; \theta_0, K_{\mu})$ is evaluated by
\begin{equation}
L(X, \mu) = \int_{t_0}^T \ell(\theta(t; \theta_0, K_{\mu}(X))) \ dt,
\label{eq:loss}
\end{equation}
where $\ell : \mathbb{T}^n \to \mathbb{R}$ is a differentiable instantaneous loss function, $t_0$ is a burn-in time, and $K_{\mu}(X)$ is shorthand for the data-dependent matrix of couplings. It will sometimes be desirable to constrain this matrix to lie in some feasible set by restricting $\mu \in \mathcal{M}$. The coupling function, the equation of motion for $\theta$ and the instantaneous loss function, $\ell$, are all differentiable in $\mu$ so we may approximate
\begin{equation}
\mu^* = \argmin_{\mu \in \mathcal{M}} \EX\limits_{X\sim D^n}\left[L(X; \mu) \right]
\label{eq:opt}
\end{equation}
by gradient descent, where $D^n$ denotes the joint distribution over the batch of node features. 

Depending on which type of dynamics we wish to encourage, we will employ two different loss functions. To encourage global synchrony (see Section~ Results~\ref{sec:results_gs}), we can simply minimize the circular variance, $\ell(\theta) = V(\theta)$. To encourage cluster synchrony, we must first assume there exists a function $Y : \mathcal{X} \to \{1,\ldots, k\}$, mapping from the node feature space to a set of discrete labels. This labeling associates to the $i^{\text{th}}$ oscillator a target cluster via the label of its feature, $l = Y(x^{(i)})$. Our goal is then to ensure the phases of all oscillators within a cluster are synchronized and all phases between clusters are maximally desynchronized. We therefore use an instantaneous loss function of the form
\begin{equation}
    \ell(\theta) = \frac{1}{2}\left(\frac{1}{k}\sum_{l=1}^k  V_l(\theta) + S(\theta)\right),
    \label{eq:cluster}
\end{equation}
where $V_l$ is the circular variance of the $l^{\text{th}}$ group, and $S$ is another loss designed to encourage splayness~\citep{Strogatz1993} among target groups in the sense that the mean phases within the groups equidistribute the unit circle. Let $\langle \theta \rangle_l$ be the average phase of oscillators having features $x_j$ such that $Y(x_j) = l$ . Then, we set
\begin{equation}
    S = \sum_{g=1}^{\floor{k/2}} \frac{1}{2g^2 k} \left| \sum_{l=1}^{k}e^{ig\langle \theta \rangle_l} \right|^2,
    \label{eq:splay}
\end{equation}
which is minimized exactly when $\{\langle \theta \rangle_l\}_{l=1}^k$ are splay \citep{Sepulchre2008}. The sum in Eq. \ref{eq:splay} is a generalization of the global order parameter, $R$, to higher circular moments. These moments pluralize the notion of synchrony so that phases are considered synchronous and maximize the $g^{\text{th}}$ moment as long as they all point in the same direction modulo $2 \pi / g$. Conversely, the sum is minimized when the phases are out of phase modulo $2 \pi / g$; i.e., they are splay. More details can be found in \citep{Sepulchre2008,Sepulchre2008b}. Note that all losses reported below are all normalized to be between 0 and 1. For the cluster synchrony experiments, the node feature space, $\mathcal{X}$, only controls the labels of oscillators and we assume that $f = g = h = \delta(0)$, so that the oscillators are equivalent to XY spins \citep[Section IV.B.1]{Acebron2005}. 

\section{Results}
\label{sec:results}
\subsection{Global synchrony} 
\label{sec:results_gs}
\begin{figure*}[t]
    \centering
    \includegraphics[width=\textwidth]{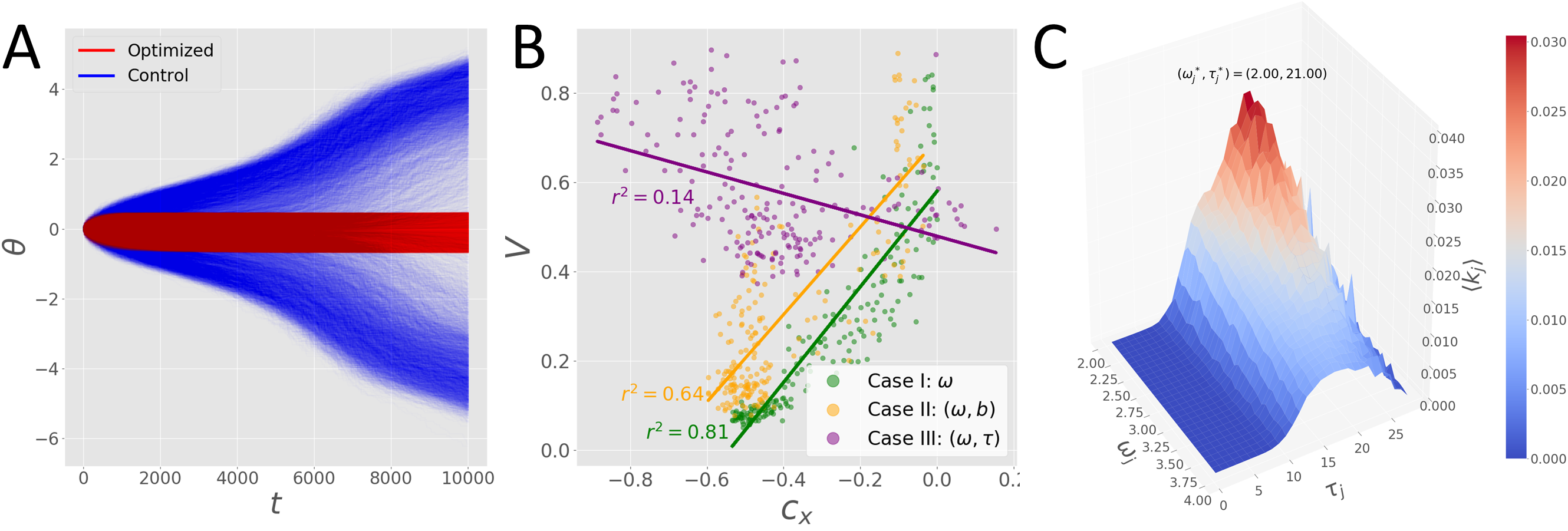}
    \caption{\emph{KuraNet learns to synchronize three types of Kuramoto models.} \emph{(A)}. The procession 10,000 of phases through time in the Size+Data Testing regime for a trained (red) versus an untrained (blue) system sharing the same average weighted network degree. Case I is shown. The untrained system desynchronizes quickly, while the optimized model remains tightly synchronized. \emph{(B)}. For Case I (green) and Case II (blue), anti-correlation of network features strongly predicts (linear at $r^2 = 0.81$, $r^2=0.64$) low circular variance, $V$. No strong relationship was found for Case III (purple). \emph{(C)}. Instead the system prioritized units with low intrinsic frequencies (long period) and time-delays on the right tail of $h$, presumably to provide a longer time-scale for potential synchronization for a larger number of strongly delayed oscillators.}
    \label{fig:brede}
\end{figure*}

Here, we learn coupling functions, $K_{\mu}$, which minimize the circular variance, $V$, of Kuramoto phases for three standard forms of Eq.~\ref{eq:km} depending on which parameters are disordered: Case I, $f = U([-1,1])$, $g = h = \delta(0)$; Case II,  $f = g = U([-1,1])$, $h = \delta(0)$; and Case III, $f = U([2,4])$, $g=\delta(0)$, $h=NB(.5,15)$, where $U$ and $NB$ are the uniform and negative binomial distributions, respectively. The discrete samples from the negative binomial distribution were used to index the time-steps of an Euler update used to solve Eq.~\ref{eq:km}. Note that Case III uses a decentered natural frequency distribution since the presence of time delays introduces a nontrivial dependence between system behavior and the natural frequency mean~\citep{Zanette2000}. In all cases, we forced $K_{\mu}(X)$ to be symmetric and positive.

To make the optimization non-trivial, we follow earlier work by \cite{Brede2008} by restricting the underlying weighted graph associated with $K_{\mu}(X)$ to have a fixed, small average degree, in this case $\langle d \rangle = \frac{1}{n}\sum_{ij}K_{\mu}( x^{(i)}, x^{(j)}) = 1$. This forces KuraNet to be judicious about its limited coupling resources: only those oscillators whose intrinsic parameters are crucial for synchrony deserve to be strongly coupled. 

To emphasize KuraNet's ability to generalize across both data and scale, we trained the system on relatively small networks ($n = 100$) and then generalized to both new data and larger network sizes ($n=10,000$). This gives rise to three testing regimes: Data Testing, where the model is exposed to new node features from the testing set for networks of size $n=100$; Size Testing, where the system is exposed to a single network of size $n=10,000$ comprising the whole \emph{training} set; and Data+Size Testing, where the model is exposed to a single network of size $n=10,000$ comprising the whole \emph{testing} set. These testing regimes measure the degree to which KuraNet can abstract beyond the observed training data and beyond finite-size effects. Each case was optimized for 10 random seeds and the best values are shown here. Further training details are found in the SI Methods section. 

\begin{table}
\centering
\footnotesize
\begin{tabular}{lllll}
Disorder  & Data  & Size & Data + Size\\
\midrule
Case I ($\omega$) & \textbf{0.0667} (0.7412) &\textbf{0.0681} (0.9172) & \textbf{0.0682} (0.9144)\\
Case II ($\omega, b$)  & \textbf{0.1527} (0.7370) & \textbf{0.1823} (0.8969) & \textbf{0.1890} (0.9378)\\
Case III ($\omega, \tau$) & \textbf{0.3800} (0.8920) & \textbf{0.7026} (0.9707) & \textbf{0.6855} (0.9753)\\
\bottomrule
\end{tabular}
\caption{Global synchrony test results measured in circular variance, $0 \leq V \leq 1$, where $V=0$ implies global synchrony and $V=1$ represents complete desynchrony. Control values in parentheses.}\label{table:global}
\end{table}
The circular variance, $V$, achieved in various testing regimes are collected in Table~\ref{table:global}. Parenthetical values show the corresponding results for an untrained control system, which nevertheless has the same average degree $\langle d \rangle$ as the trained one. For all three cases and in all three testing regimes, the optimized model performs substantially better than the control model. The most striking results are achieved in Case I, where the optimized system achieves nearly perfect synchrony in all three testing regimes (Figure~\ref{fig:brede} A).

\begin{figure*}[t]
\centering
\includegraphics[width=0.85\textwidth]{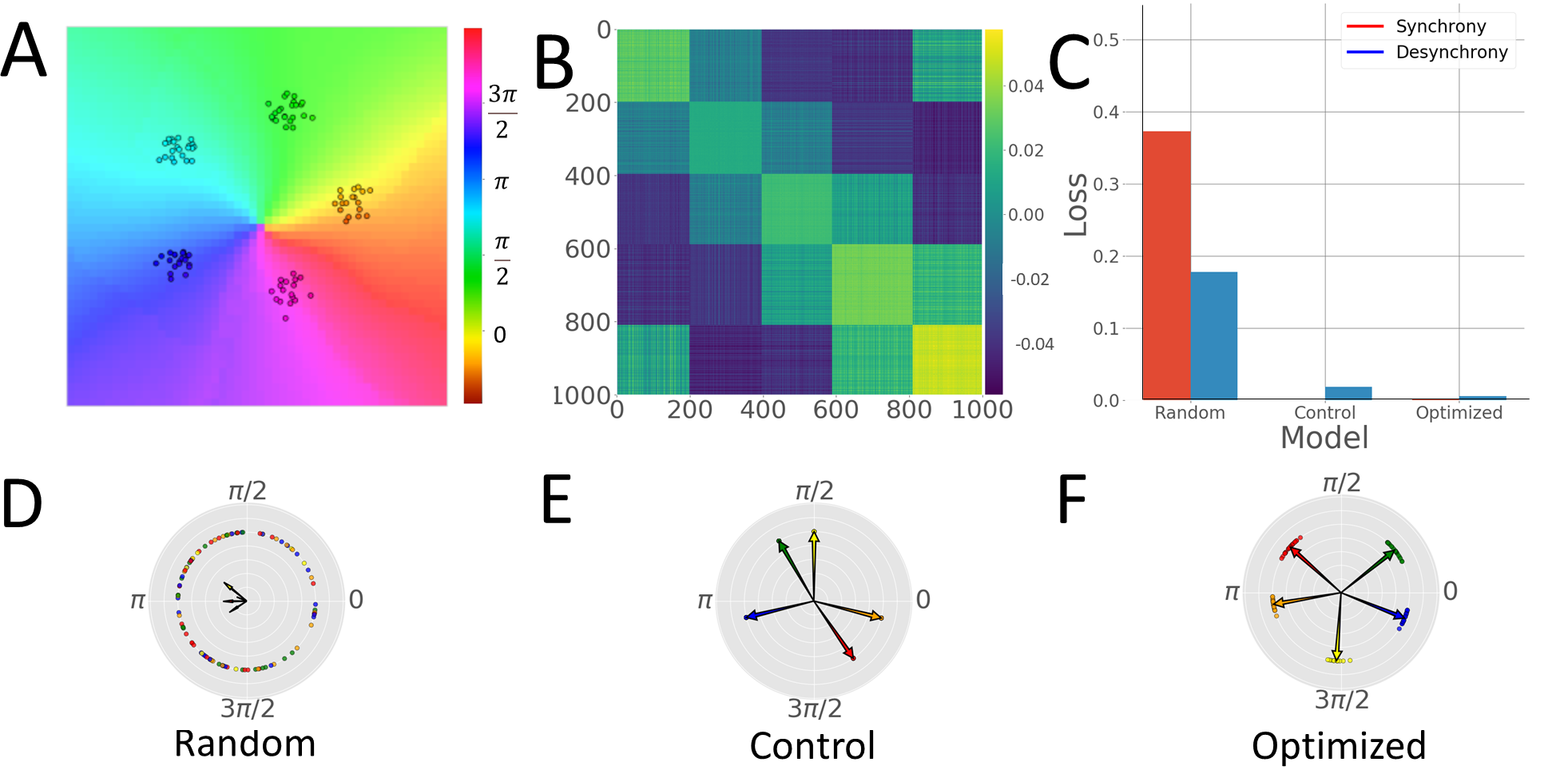}
\caption{\emph{Cluster synchrony for a 5-component gaussian mixture model.} \emph{(A)} The phase landscape in $\mathcal{X} = \mathbb{R}^2$ is plotted by recording the asymptotic phase of a single oscillator, $i$, in a fixed batch (small circles) as its 2-D feature was varied over a grid of 100 x 100 values. \emph{(B)} The trained system learns to produce coupling matrices with a clear block-diffusive structure whose cells correspond to the true labeling function on $\mathcal{X}$. \emph{(C)} However, not just any block coupling structure can produce this effect, since the optimized model (right bars) achieves a lower total loss than both a randomized model (left bars) and a block control model (middle bars) with positive intra-cluster couplings and negative inter-cluster couplings. \emph{(D,E,F)} The differences among these models is all the clearer when plotting terminal phases (small dots) and cluster mean fields (colored arrows). Here, phases and mean fields are colored according to ground truth clusters, not angle. While the control model can easily exhibit strong intra-group synchrony \emph{(E}, only the balance of positive and negative weights in the learned diffusive couplings of the optimized model \emph{(F)} can also produce reliable splay configurations.}\label{fig:cluster}
\end{figure*}

To help explain the results of our optimization, we introduce a generalized version of the graph-induced autocorrelation function proposed by \cite{Brede2008},
\begin{equation}
    \label{eq:bq}
    C_x = \frac{\sum_{i,j}K_{\mu}( x^{(i)}, x^{(j)}) \langle x^{(i)} - \langle x \rangle, x^{(j)} - \langle x \rangle \rangle}{\sum_{i,j}K_{\mu}( x^{(i)}, x^{(j)})(x^{(i)} - \langle x \rangle)^2}, 
\end{equation}
where $\langle x \rangle = \frac{1}{n} \sum_{i=1}^n x^{(i)} \in \mathcal{X}$. Eq.~\ref{eq:bq} measures the correlation between random node features with respect to the couplings, $K_{\mu}( x^{(i)}, x^{(j)})$. Plotting $V$ versus $C_x$ for each of the three optimizations reveals that graph-induced anti-correlation between nodes is strongly associated with synchrony (linear at $r^2 = 0.81, r^2 = 0.64$) in Case I and Case II, but not Case III ($r^2=0.14$). Hence, synchrony in the former cases is promoted when the limited coupling resource is spent between nodes with disparate features, effectively reducing the heterogeneity of the system. A similar result was found by \cite{Brede2008}, but only for the equivalent of our Case I. 

Case III is evidently rather different, but we can gain some insight by plotting the learned coupling function. Discretizing the sample space of $\omega$ and $\tau$ into 30 bins and plotting the average fan-in weight $\langle k_j \rangle =  \frac{1}{n}\sum_{i}K_{\mu}( x^{(i)}, x^{(j)})$ associated to each bin over 100 batches shows a marked peak at $(\omega^*, \tau^*) = (2.00, 21.00)$. Notably, the peak occurs at the lowest intrinsic frequency and at a time delay slightly to the right of the mode of $h$ ($\tau = 15$). This was consistent for all of the best random seeds for this case. The mean of the intrinsic frequency distribution sets an effective time-scale for the network and it is known that delays which are long compared to this time-scale tend to impede synchrony~\citep{Zanette2000}. Our result suggests that this impediment can be overcome when weights simultaneously emphasize the longest time-scale (i.e. lowest intrinsic frequencies) and those delays which are as long as possible without being too far from the mode of $h$. We simply take the results from Case III as a computational hypothesis about disordered time delays in the Kuramoto model, which we leave for future investigation.

\subsection{Cluster synchrony}
\label{sec:results_ss}
\begin{figure*}[t]
\centering
\includegraphics[width=.85\textwidth]{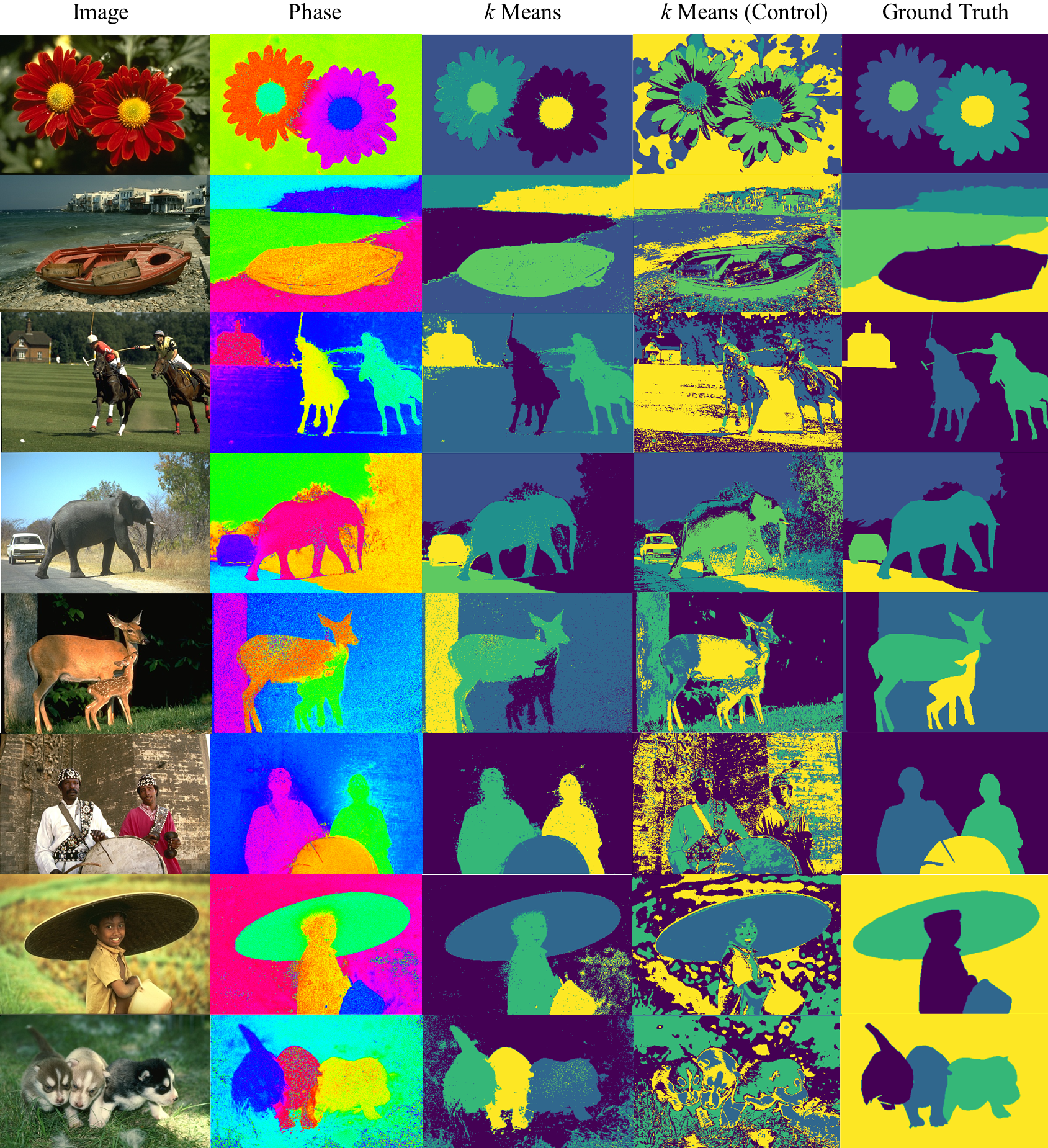}
\caption{\emph{Image segmentation results.} KuraNet is trained on ground truth segmentation masks and pixel features (locations and channel intensities) from half the pixels in an image. Here, we show the learned segmentation ability on the whole image (train and test; test only results averaged in the last row of Table~\ref{table:cluster}). \emph{(First column)} Original image. \emph{(Second column)} Limiting phase of KuraNet on these images after training (hue represents phase). \emph{(Third column)} \emph{k}-means clustering is used to discretize the phase into segments. \emph{(Fourth column)} Corresponding $k$-means segmentations using raw pixels as data instead of KuraNet-produced phases. \emph{(Fifth column)} Ground truth segmentation. Note that the correspondence between predicted segments and ground truth segments does not rely on the absolute labeling of either segmentation (here, the color), but rather the existence of a bijection between the two sets of labels which preserves the relationship between any two pixels' labels (i.e., same vs. different) (see SI Methods.)}\label{fig:images}
\end{figure*}
Next, we turn to the case of cluster synchrony, in which oscillators split into internally coherent factions. Cluster synchrony has attracted a great deal of recent interest and has been observed, for example, in flocking animals~\citep{Ramos2019}, models of visual processing~\citep{Reichert2013}, and power grids~\citep{Pecora2014}. Here, we are interested in the stronger condition in which oscillators not only synchronize into clusters but also where these clusters are maximally separated in the sense that the cluster phase averages, $\langle \theta\rangle_l$, are equally spread out across the unit circle. We investigated optimized cluster synchrony for two types of data: simple mixture distributions on the plane, and images taken from the Berkeley Segmentation Data Set (BSDS)~\citep{MartinFTM01}.

\begin{table}
\footnotesize
\centering
\begin{tabular}{lllll}
Dist.  & Data  & Size & Data + Size\\
\midrule
Moons  & \textbf{0.0154} (0.5000) &\textbf{0.0012} (0.5000) & \textbf{0.0011} (0.5000)\\
Circles & \textbf{0.0257} (0.5000) &\textbf{0.0015} (0.5000) & \textbf{0.0021} (0.5000)\\
Spirals & \textbf{0.0182} (0.5000) &\textbf{0.0096} (0.5000) & \textbf{0.0094} (0.5000)\\
GMM\emph{k} (avg.) & \textbf{0.0144} (0.4904) &\textbf{0.0069} (0.5424) & \textbf{0.0073} (0.5711)\\
Images (avg.) & \textbf{0.0376} (0.7464) &\textbf{0.0088} (0.8363) & \textbf{0.0093} (0.8360)\\
\bottomrule
\end{tabular}
\caption{Cluster synchrony test results, measured in the averaged synchrony and splay losses, $0 \leq L \leq 1$, where $L = 0$ implies full intra-group synchrony and inter-group splayness and $L = 1$ implies the opposite. Control values in parentheses. Full Gaussian mixture and image results are found in the SI Detailed Clustering Results.}
\label{table:cluster}
\end{table}

For simple mixture distributions, we used several standard distributions from Python's \texttt{sklearn} software: two interlinked crescent moons (``Moons''), two concentric circles (``Circles''), two interlinked spirals (``Spirals''),  as well as nine Gaussian mixture models (GMM$k$) with between $k = 2, \ldots, 10$ means spaced evenly around a ring of radius 10 (results here only show $k=5$, but more are available in the SI Detailed Clustering Results). Node features were the coordinates $(p,q)$ of the corresponding datum and labels came from the labels of mixture components. 

As before, we are interested in both generalization to new data and to larger networks, resulting in the three testing regimes described above. Mixture distributions used training and testing network sizes of $n=100$ and $n=10,000$ as before. The coupling function architecture was identical to that of Section~\ref{sec:results_ss}, except that we now permit negative couplings. The coupling matrix was still normalized so that each unit had an (absolute) in-degree of 1.0 before being symmetrized by $K \leftarrow .5 (K + K^T)$. For details, see SI Methods. 

After training, we found that the model was able to generalize to correct splay dynamics across all testing regimes far better than an untrained model with random parameters but with the same degree normalization (Table~\ref{table:cluster}). A fuller picture of the trained system is provided by plotting the limiting phase of a single oscillator (five-class Gaussian mixture shown in Figure~\ref{fig:cluster}A) as its feature vector is varied across a grid of values. For two cluster data (Moons, Circles, Spirals), the randomized model tended to synchronize globally for a loss of .5, while the optimized model achieves a loss of nearly 0. The optimized system also achieved nearly perfect cluster synchrony in the multi-class cases (e.g. GMM5), where the random model's partial synchrony only produced a loss of about an order of magnitude worse. Videos of the dynamics can be found in the SI.

Plotting couplings sorted by cluster index for a sample of 1,000 training oscillators revealed that the system had learned in all cases a distinctive mixture of block and diffusive connectivity (Figure~\ref{fig:cluster}B). Couplings between oscillators in different groups were uniform as a function of the group indices, and these couplings grew weaker and more negative as the group features grew more distant in the plane. Earlier work~\citep{Wang2005, Schaub2016, Tiberi2018} has implicated one of these properties, block coupling, in the emergence of cluster synchrony, although these studies did not include our extra condition of mutual desynchrony among clusters. To compare our optimized couplings to coupling strategies in this earlier work, we examined an additional control model with block couplings normalized exactly as in the optimized case but with a fixed intra-group (inter-group) coupling of $b > 0$ ($b < 0$). The value of $b$ was chosen to such that the optimized and control settings would have the same maximum absolute coupling strength, $\max\limits_{i,j }|K_{ij}|$, so that attraction and repulsion between groups would be on roughly the same scale in both cases. We found the block control could successfully synchronize each cluster far better than the random setting (Figure~\ref{fig:cluster}C), although the independence of clusters prevented any rigid splay behavior, particularly for $k>2$. This is especially evident when the mean fields (Figure~\ref{fig:cluster} D-F) are plotted for the true clusters in each condition. This result suggests that the learned diffusivity in the optimized system enforces cluster separation by tuning the balance between the relatively few intra-group interactions and relatively many inter-group interactions which prevail in the case of $k>2$.

For the image experiments, we chose eight images from BSDS. Here, KuraNet functions as a form of image segmentation (Fig.~\ref{fig:images}). Various other oscillator-based models of image segmentation have been proposed \citep{Yu2009, Meier2014},  though, to our knowledge, the current work is the first end-to-end differentiable model trainable by gradient descent. Similar to the mixture distribution case, node features include the pixel location location $(p,q)$, although we now also include the the RGB channel intensities $I_r, I_g, I_b\in [0,255]$ at the location $(p,q)$. Labels came from the ground truth segments of the BSDS data set. Training was identical to the mixture distribution case except that training and test sets were considerably larger owing to the image sizes ($\sim 150,000$ pixels). Note that KuraNet learns to segment one image at a time (as opposed to multiple images). We speculate on ways to achieve this larger goal in Section \ref{sec:discussion}.

Performance on images was favorable (Table~\ref{table:cluster}, last row; full results, SI Detailed Clustering Results), and Fig. \ref{fig:images} shows the results on the full images (training plus test sets). Note that the images in Figure \ref{fig:images} represent the dynamics on very large networks (again, $\sim 150,000$ nodes) even though KuraNet was only ever exposed to dynamics on networks three orders of magnitude smaller. Discretizing the limiting phase configuration (Fig.~\ref{fig:images}, second column) using $k$-means clustering (Fig.~\ref{fig:images}, third column) and measuring performance using the standard symmetric best dice score $0 \leq SBD \leq 1$ \citep{Scharr2016} shows that KuraNet achieves an accuracy of $.7001$ vs a value of $.3719$ (slightly above 
chance level) attained by $k$-means on raw pixel intensities (fourth column). These results cast KuraNet as an interesting new direction for image processing using complex systems.

\section{Discussion}
\label{sec:discussion}

Recent work at the intersection of machine learning and dynamical systems has typically taken a somewhat model-free approach, estimating couplings~\citep{Nitzan2017} or learning dynamics~\citep{Chen2018} while assuming little about the underlying equation of motion. Here, we demonstrate the value of the model-based setting, whereby the parameter space of a given dynamics of interest, Kuramoto flows expressible by Eq.~\ref{eq:km}, is systematically explored. We believe this type of model-based statistical approach to studying specific physical or biological models holds great promise. 

The ability of KuraNet to generate computational hypotheses about collective motion in particularly complicated dynamical systems (Sec. Results~\ref{sec:results_gs}) and to learn data-driven cluster synchrony patterns (Sec. Results~\ref{sec:results_ss}), represents only the early steps of this model-based approach. For example, note that our pairwise parallel processing of nodes relies on the fact that node features are statistically independent. Future work could relax this assumption and introduce more sophisticated coupling functions capable of learning more complicated relations among nodes in line with current work on complex network generation~\citep{Liao2019}. Similarly, future image segmentation work could use more hierarchical network structures like those in the standard convolutional neural networks~\citep{Krizhevsky2012} and in earlier oscillatory models~\citep{Yu2009}. Hierarchical networks whose couplings represent invariant features of natural images are likely necessary for training an oscillatory model to synchronize multiple images across a data set (as opposed to one at a time). Furthermore, our method could be applied to dynamical regimes beyond synchrony, for example the elusive glassiness speculated to appear in some oscillator networks~\citep{Iatsenko2014a,Ottino2018}.

Importantly, the framework we have called ``KuraNet'' can be easily applied beyond its namesake if Eq.~\ref{eq:km} is replaced with another dynamics of interest. In that light, the proposed method becomes a way of casting the more general problem of understanding collective motion under conditions of quenched disorder in the framework of statistical learning. For example, the outlined technique could be used to learn coupling functions in neural systems or epidemiological networks, where disorder takes the form of heterogeneous cell biophysics \citep{Meijas2014} and infection properties \citep{Scata2016}, respectively. These applications will have to adapt the basic KuraNet approach to domain idiosyncracies, and we are intrigued by how this adaptation may give rise to insights in both the natural sciences and machine learning.

\section{Acknowledgments}
This work was supported by ONR (N00014-19-1-2029) and NSF (IIS-1912280). Additional support was provided by the ANR-3IA Artificial and Natural Intelligence Toulouse Institute (ANR-19-PI3A-0004).

\begin{thebibliography}{10}

\bibitem{Ursey1999}
W.~Martin Usrey and R.~Clay Reid.
\newblock Synchronous activity in the visual system.
\newblock {\em Annual Review of Physiology}, 61(1):435--456, 1999.
\newblock PMID: 10099696.

\bibitem{Salinas2001}
Emilio Salinas and Terrence~J. Sejnowski.
\newblock {Correlated neuronal activity and the flow of neural information}.
\newblock {\em Nature Reviews Neuroscience}, 2(8):539--550, 2001.

\bibitem{He2003}
Daihai He and Lewi Stone.
\newblock {Spatio-temporal synchronization of recurrent epidemics}.
\newblock {\em Proceedings of the Royal Society B: Biological Sciences},
  270(1523):1519--1526, 2003.

\bibitem{Strogatz1993}
Steven~H. Strogatz and Renato~E. Mirollo.
\newblock {Splay states in globally coupled Josephson arrays: Analytical
  prediction of Floquet multipliers}.
\newblock {\em Physical Review E}, 47(1):220--227, 1993.

\bibitem{Vasudevan2015}
K.~Vasudevan, M.~Cavers, and A.~Ware.
\newblock Earthquake sequencing: chimera states with kuramoto model dynamics on
  directed graphs.
\newblock {\em Nonlinear Processes in Geophysics}, 22(5):499--512, 2015.

\bibitem{Rodrigues2016}
Francisco~A. Rodrigues, Thomas K.~DM. Peron, Peng Ji, and J\"{u}rgen Kurths.
\newblock The kuramoto model in complex networks.
\newblock {\em Physics Reports}, 610:1?98, Jan 2016.

\bibitem{Knoblauch2012}
Andreas Knoblauch, Florian Hauser, Marc~Oliver Gewaltig, Edgar K{\"{o}}rner,
  and G{\"{u}}nther Palm.
\newblock {Does spike-timing-dependent synaptic plasticity couple or decouple
  neurons firing in synchrony?}
\newblock {\em Frontiers in Computational Neuroscience}, 6(AUGUST), 2012.

\bibitem{Ramos2019}
Rita~Parada Ramos, Sancho~Moura Oliveira, Susana~Margarida Vieira, and
  Anders~Lyhne Christensen.
\newblock {Evolving flocking in embodied agents based on local and global
  application of Reynolds' rules}.
\newblock {\em PLoS ONE}, 14(10):1--16, 2019.

\bibitem{Hansen2020}
Jakob Hansen and Robert Ghrist.
\newblock Opinion dynamics on discourse sheaves, 2020.

\bibitem{Chen2018}
Ricky~T.Q. Chen, Yulia Rubanova, Jesse Bettencourt, and David Duvenaud.
\newblock {Neural Ordinary Differential Equations}.
\newblock In {\em 32nd Conference on Neural Information Processing Systems
  (NeurIPS 2018)}, Montr{\'{e}}al, Canada, 2018. Curran Associates.

\bibitem{Champion2019}
Kathleen Champion, Bethany Lusch, J.~{Nathan Kutz}, and Steven~L. Brunton.
\newblock {Data-driven discovery of coordinates and governing equations}.
\newblock {\em Proceedings of the National Academy of Sciences of the United
  States of America}, 116(45):22445--22451, 2019.

\bibitem{chalvidal2021}
Mathieu Chalvidal, Matthew Ricci, Rufin VanRullen, and Thomas Serre.
\newblock Go with the flow: Adaptive control for neural odes.
\newblock In {\em International Conference on Learning Representations}, 2021.

\bibitem{Brede2008}
Markus Brede.
\newblock {Synchrony-optimized networks of non-identical Kuramoto oscillators}.
\newblock {\em Physics Letters, Section A: General, Atomic and Solid State
  Physics}, 372(15):2618--2622, 2008.

\bibitem{Kuramoto1975}
Yoshiki Kuramoto.
\newblock {Self-Entrainment of a Population of Coupled Non-Linear Oscillators}.
\newblock In H~Araki, editor, {\em 39th International Symposium on Mathematical
  Problems in Theoretical Physics}, New York, 1975. Springer-Verlag.

\bibitem{Strogatz2000}
Steven~H. Strogatz.
\newblock From kuramoto to crawford: exploring the onset of synchronization in
  populations of coupled oscillators.
\newblock {\em Physica D: Nonlinear Phenomena}, 143(1):1--20, 2000.

\bibitem{Wang2005}
Wei Wang and Jean Jacques~E. Slotine.
\newblock {On partial contraction analysis for coupled nonlinear oscillators}.
\newblock {\em Biological Cybernetics}, 92(1):38--53, 2005.

\bibitem{Schaub2016}
Michael~T. Schaub, Neave O'Clery, Yazan~N. Billeh, Jean~Charles Delvenne,
  Renaud Lambiotte, and Mauricio Barahona.
\newblock {Graph partitions and cluster synchronization in networks of
  oscillators}.
\newblock {\em Chaos}, 26(9), 2016.

\bibitem{Tiberi2018}
Lorenzo Tiberi, Chiara Favaretto, Mario Innocenti, Danielle~S. Bassett, and
  Fabio Pasqualetti.
\newblock {Synchronization patterns in networks of Kuramoto oscillators: A
  geometric approach for analysis and control}.
\newblock {\em 2017 IEEE 56th Annual Conference on Decision and Control, CDC
  2017}, 2018-January:481--486, 2018.

\bibitem{Malkin1949}
I.G. Malkin.
\newblock {\em {The methods of Lyapunov and Poincare in the theory of nonlinear
  oscillations}}.
\newblock Editorial URSS, Moscow, 1949.

\bibitem{Winfree1967}
Arthur~T. Winfree.
\newblock Biological rhythms and the behavior of populations of coupled
  oscillators.
\newblock {\em Journal of Theoretical Biology}, 16(1):15--42, 1967.

\bibitem{Acebron2005}
Juan~A. Acebr\'on, L.~L. Bonilla, Conrad~J. P\'erez~Vicente, F\'elix Ritort,
  and Renato Spigler.
\newblock The kuramoto model: A simple paradigm for synchronization phenomena.
\newblock {\em Rev. Mod. Phys.}, 77:137--185, Apr 2005.

\bibitem{Gupta2014}
Shamik Gupta, Alessandro Campa, and Stefano Ruffo.
\newblock {Kuramoto model of synchronization: Equilibrium and nonequilibrium
  aspects}.
\newblock {\em Journal of Statistical Mechanics: Theory and Experiment},
  2014(8), 2014.

\bibitem{Iatsenko2014a}
D.~Iatsenko, P.~V.E. McClintock, and A.~Stefanovska.
\newblock {Glassy states and super-relaxation in populations of coupled phase
  oscillators}.
\newblock {\em Nature Communications}, 5(May):1--11, 2014.

\bibitem{Ottino2018}
Bertrand Ottino-L\"{o}ffler and Steven~H. Strogatz.
\newblock Volcano transition in a solvable model of frustrated oscillators.
\newblock {\em Physical Review Letters}, 120(26), Jun 2018.

\bibitem{Skardal2018}
Per~Sebastian Skardal.
\newblock {Stability Diagram, Hysteresis, and Critical Time Delay and Frequency
  for the Kuramoto Model with Heterogeneous Interaction Delays}.
\newblock {\em International Journal of Bifurcation and Chaos}, 28(5):1--15,
  2018.

\bibitem{Olmi2014}
Simona Olmi, Adrian Navas, Stefano Boccaletti, and Alessandro Torcini.
\newblock {Hysteretic transitions in the Kuramoto model with inertia}.
\newblock {\em Physical Review E - Statistical, Nonlinear, and Soft Matter
  Physics}, 90(4):1--34, 2014.

\bibitem{Gleiser2006}
P.~M. Gleiser and D.~H. Zanette.
\newblock {Synchronization and structure in an adaptive oscillator network}.
\newblock {\em European Physical Journal B}, 53(2):233--238, 2006.

\bibitem{Kelly2011}
David Kelly and Georg~A. Gottwald.
\newblock {On the topology of synchrony optimized networks of a Kuramoto-model
  with non-identical oscillators}.
\newblock {\em Chaos}, 21(2), 2011.

\bibitem{Nitzan2017}
Mor Nitzan, Jose Casadiego, and Marc Timme.
\newblock {Revealing physical interaction networks from statistics of
  collective dynamics}.
\newblock {\em Science Advances}, 3(2):2--8, 2017.

\bibitem{Sepulchre2008}
Rodolphe Sepulchre, Derek~A. Paley, and Naomi~Ehrich Leonard.
\newblock {Stabilization of planar collective motion with limited
  communication}.
\newblock {\em IEEE Transactions on Automatic Control}, 53(3):706--719, 2008.

\bibitem{Sepulchre2008b}
Rodolphe Sepulchre, Derek~A. Paley, and Naomi~Ehrich Leonard.
\newblock {Stabilization of planar collective motion with limited
  communication}.
\newblock {\em IEEE Transactions on Automatic Control}, 53(3):706--719, 2008.

\bibitem{Zanette2000}
Dami{\'{a}}n~H. Zanette.
\newblock {Propagating structures in globally coupled systems with time
  delays}.
\newblock {\em Physical Review E - Statistical Physics, Plasmas, Fluids, and
  Related Interdisciplinary Topics}, 62(3 A):3167--3172, 2000.

\bibitem{Reichert2013}
David~P. Reichert and Thomas Serre.
\newblock Neuronal synchrony in complex-valued deep networks, 2014.

\bibitem{Pecora2014}
Louis~M. Pecora, Francesco Sorrentino, Aaron~M. Hagerstrom, Thomas~E. Murphy,
  and Rajarshi Roy.
\newblock {Cluster synchronization and isolated desynchronization in complex
  networks with symmetries}.
\newblock {\em Nature Communications}, 5(May), 2014.

\bibitem{MartinFTM01}
D.~Martin, C.~Fowlkes, D.~Tal, and J.~Malik.
\newblock A database of human segmented natural images and its application to
  evaluating segmentation algorithms and measuring ecological statistics.
\newblock In {\em Proc. 8th Int'l Conf. Computer Vision}, volume~2, pages
  416--423, July 2001.

\bibitem{Yu2009}
Guoshen Yu and Jean~Jacques Slotine.
\newblock {Visual grouping by neural oscillator networks}.
\newblock {\em IEEE Transactions on Neural Networks}, 20(12):1871--1884, 2009.

\bibitem{Meier2014}
Martin Meier, Robert Haschke, and Helge~J Ritter.
\newblock {Perceptual grouping by entrainment in coupled kuramoto oscillator
  networks}.
\newblock {\em Network: Computation in Neural Systems}, 25(June):72--84, 2014.

\bibitem{Scharr2016}
Hanno Scharr, Massimo Minervini, Andrew~P. French, Christian Klukas, David~M.
  Kramer, Xiaoming Liu, Imanol Luengo, Jean~Michel Pape, Gerrit Polder,
  Danijela Vukadinovic, Xi~Yin, and Sotirios~A. Tsaftaris.
\newblock {Leaf segmentation in plant phenotyping: a collation study}.
\newblock {\em Machine Vision and Applications}, 27(4):585--606, 2016.

\bibitem{Liao2019}
Renjie Liao, Yujia Li, Yang Song, Shenlong Wang, Charlie Nash, William~L.
  Hamilton, David Duvenaud, Raquel Urtasun, and Richard~S. Zemel.
\newblock {Efficient Graph Generation with Graph Recurrent Attention Networks}.
\newblock In {\em 33rd Conference on Neural Information Processing Systems
  (NeurIPS 2019)}, Vancouver, Canada, 2019. Curran Associates.

\bibitem{Krizhevsky2012}
Alex Krizhevsky, Ilya Sutskever, and Geoffrey~E. Hinton.
\newblock {ImageNet Classification with Deep Convolutional Neural Networks}.
\newblock In {\em NIPS'12 Proceedings of the 25th International Conference on
  Neural Information Processing Systems}, pages 1097--1105, Lake Tahoe, Nevada,
  2012. Curran Associates.

\bibitem{Meijas2014}
Jorge~F. Mejias and Andr\'{e} Longtin.
\newblock Differential effects of excitatory and inhibitory heterogeneity on
  the gain and asynchronous state of sparse cortical networks.
\newblock {\em Frontiers in Computational Neuroscience}, 8:107, 2014.

\bibitem{Scata2016}
Marialisa Scat{\`{a}}, Alessandro {Di Stefano}, Pietro Li{\`{o}}, and Aurelio
  {La Corte}.
\newblock {The Impact of Heterogeneity and Awareness in Modeling Epidemic
  Spreading on Multiplex Networks}.
\newblock {\em Scientific Reports}, 6(August):1--13, 2016.

\end{thebibliography}

\end{document}